\RequirePackage{fix-cm}
\documentclass[twocolumn,epjc3]{svjour3}
\smartqed  
\RequirePackage{graphicx}

\usepackage{amsmath}
\usepackage{xcolor}
\DeclareMathOperator{\erf}{erf}
\journalname{Eur. Phys. J. C}
\begin{document}

\title{Steering of Sub-GeV electrons by ultrashort Si and Ge bent crystals}

\author{  
A.I. Sytov \thanksref{addr1,addr2} \and L. Bandiera \thanksref{addr1,e1} \and D. De Salvador \thanksref{addr4} \and A. Mazzolari \thanksref{addr1}  \and E. Bagli \thanksref{addr1} \and A. Berra \thanksref{addr3} \and  S. Carturan \thanksref{addr4} \and C. Durighello \thanksref{addr1,addr4} \and G. Germogli \thanksref{addr1} \and V. Guidi \thanksref{addr1} \and P. Klag \thanksref{addr5} \and W. Lauth \thanksref{addr5} \and G. Maggioni \thanksref{addr4} \and M. Prest \thanksref{addr3} \and M. Romagnoni \thanksref{addr1} \and 
\\ V.V. Tikhomirov \thanksref{addr2} \and E. Vallazza \thanksref{addr6}}

\thankstext{e1}{e-mail: bandiera@fe.infn.it}


\institute{INFN Sezione di Ferrara, Dipartimento di Fisica e Scienze della Terra, Universit\`{a} di Ferrara Via Saragat 1, 44100 Ferrara, Italy \label{addr1}
           \and
           Institute for Nuclear Problems, Belarusian State University, Bobruiskaya 11, 220030 Minsk, Belarus \label{addr2}
           \and
           INFN Laboratori Nazionali di Legnaro, Viale dell'Universit\`{a} 2, 35020 Legnaro, Italy\\
           Dipartimento di Fisica, Universit\`{a} di Padova, Via Marzolo 8, 35131 Padova, Italy \label{addr4} 
           \and 
           Institut f\"{u}r Kernphysik der Universit\"{a}t Mainz,  D-55099 Mainz, Germany \label{addr5} 
           \and
           Universit\`{a} dell'Insubria, via Valleggio 11, 22100 Como, Italy and INFN Sezione di Milano Bicocca,\\
           Piazza della Scienza 3, 20126 Milano, Italy \label{addr3} 
           \and
           INFN Sezione di Trieste, Via Valerio 2, 34127 Trieste, Italy \label{addr6}
}

\date{Received: date / Accepted: date}

\maketitle

\begin{abstract}
{We report the observation of the steering of 855 MeV electrons by bent silicon and germanium crystals at the MAinzer MIkrotron. 15 $\mu $m long crystals, bent along (111) planes, were exploited to investigate orientational coherent effects. By using a piezo-actuated mechanical holder, which allowed to remotely change the crystal curvature, it was possible to study the steering capability of planar channeling and volume reflection vs. the curvature radius and the atomic number, Z. For silicon, the channeling efficiency exceeds 35 \%, a record for negatively charged particles. This was possible due to the realization of a crystal with a thickness of the order of the dechanneling length. On the other hand, for germanium the efficiency is slightly below 10 \% due to the stronger contribution of multiple scattering for a higher-Z material. Nevertheless this is the first evidence of negative beam steering by planar channeling in a Ge crystal. Having determined for the first time the dechanneling length, one may design a Ge crystal based on such knowledge providing nearly the same channeling efficiency of silicon. The presented results are relevant for crystal-based beam manipulation as well as for the generation of e.m. radiation in bent and periodically bent crystals.}

\end{abstract}

\section{Introduction}
\label{intro}
\begin{figure*}
\resizebox{85mm}{!}{\includegraphics{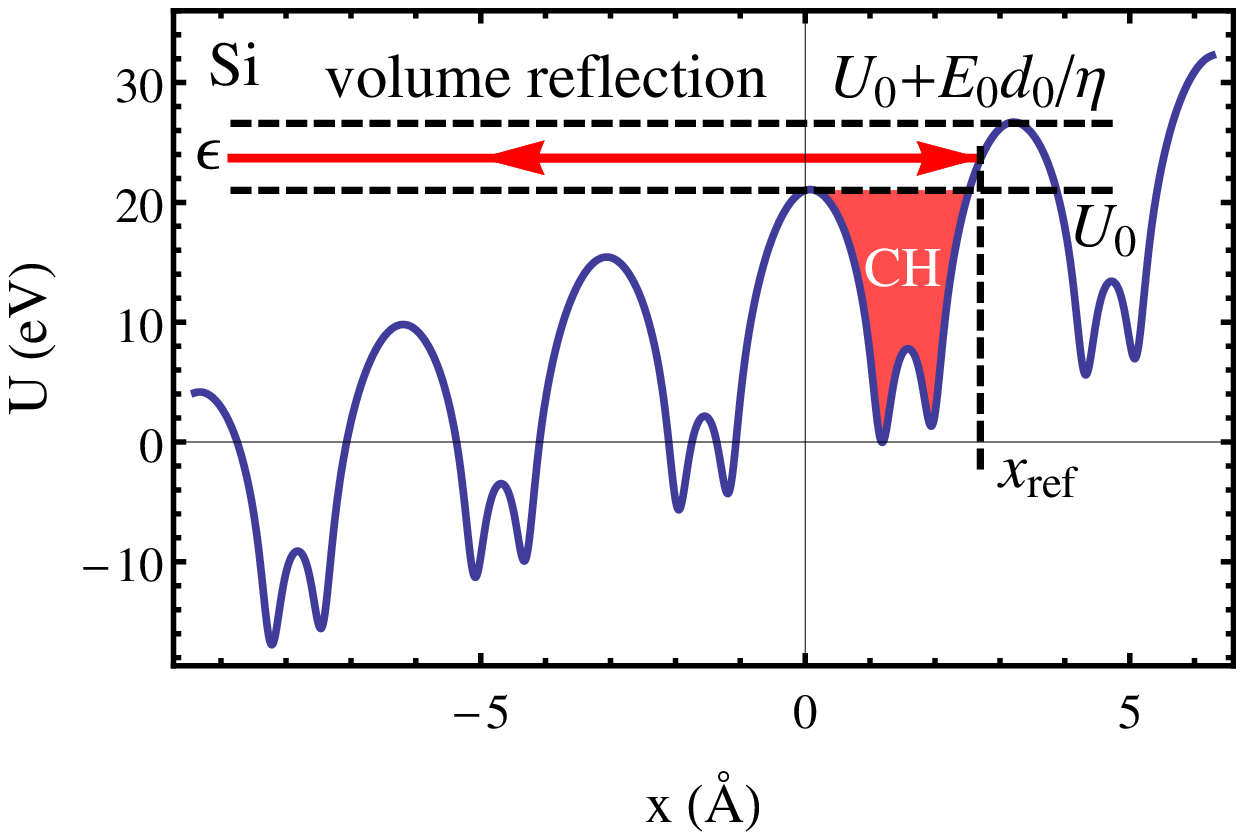}}
\resizebox{85mm}{!}{\includegraphics{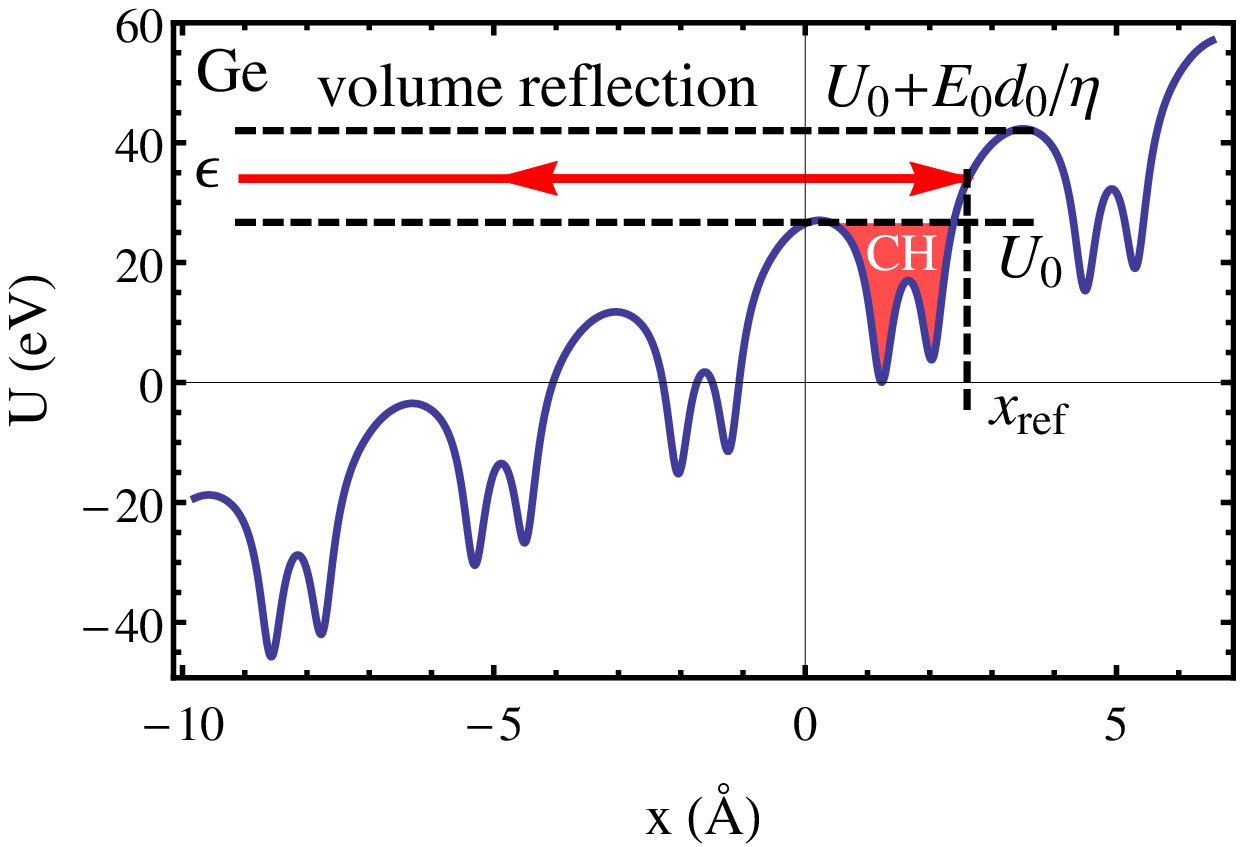}}
\caption{\label{Fig1} Interplanar potential of (111) Si and Ge crystals for maximal bending radii (4.76 cm and 1.83 cm respectively) used in the experiments.}
\end{figure*}

Beam steering based on the coherent interaction of charged particle beams with bent crystals found several applications in accelerator physics. In particular, crystal-based beam collimation and extraction were successfully investigated at several accelerator machines, such as  U70, SPS, RHIC, Tevatron and LHC \cite{U70,U702,Tevatron,Tevatron2,UA9,UA92,LHC}. The main idea of crystal-assisted beam steering, firstly proposed by Tsyganov in 1976 \cite{Tsyganov}, relies on planar channeling \cite{Lindhard}, holding a charged particle in a potential well (see Fig. \ref{Fig1}) formed by the electric field of two neighboring atomic planes. If a crystal is bent, charged beam will be steered since its trajectory is confined under channeling conditions along bent crystal planes.
Another coherent effect in a bent crystal, i.e., the so-called volume reflection (VR) \cite{VR}, consists of the reversal of charged particle momentum by the potential barrier of a bent crystal plane. 
Fig. \ref{Fig1} displays the potential energy of the bent (111) plane Si (left) and Ge (right) crystals. The figure also shows the potential well in which a channeled particle is captured (marked as \textit{CH}), while the red line with a double arrow depicts the \textit{reflection} of particle momentum under VR. Through VR, charged particles are deflected to a smaller bending angle with respect to channeling, however this method provides a considerably larger and adjustable angular acceptance, equal to the crystal curvature, and higher deflection efficiency.

Over the years, positive particle beam steering has been well investigated in a wide range of energies, from few MeV up to the recent result at LHC with 6.5 TeV protons. Conversely, the negatively charged particle case has been poorly investigated and only recently, thanks to the realization of very short bent Si crystals, it has become possible to steer negatively charged particles beams \cite{PRA2012,NIMB2013,PRL2013,PRL2015,PRL2014,EPJC2017,PRABSLAC,PRLSLAC,PS2011,SPS2009ax,SPS2009}. However, the possibility to steer electron beams is promising for applications in electron-positron collider collimation systems \cite{PRL2013,ILC,CLIC,RREPS2013,SERYI} as well as innovative high-intensity $X$- or $\gamma$-radiation sources \cite{PRA2012,NIMB2013,PRL2013,PRL2015,PS2016}. 

Usually, silicon is selected as prime material for the fabrication of bent crystals due to its high-quality crystalline lattice and low cost. Nevertheless, other materials, like germanium, which provides a higher atomic number, \textit{Z}, than silicon and can be also realized with a similar perfection, deserve investigation. Indeed, since a Ge crystal provides a stronger potential, one expects an increase in the angular acceptance for channeling and an enhancement of e.m. radiation emission. Currently, a few channeling experiments have been performed with bent Ge, only with positively charged particles and only in the hundreds GeV energy range \cite{Ge0,Ge1,Ge2,Ge3}, while with electrons and at lower energies there are no data in literature due to the technical difficulties of fabrication of an ultra-short bent Ge crystal.

In this paper, we present an investigation on sub-GeV electron steering by both silicon and germanium bent crystals under channeling and VR. With the aim of determining the different behavior of these effects vs. the atomic number \textit{Z}, two 15 $\mu $m Si and Ge crystals, bent along the (111) planes, were selected and an experiment was performed at the Mainz Mikrotron (MAMI) with 855 MeV electrons. We also investigated the dependence of the channeling efficiency and of the dechanneling length, which is the main parameter for planar channeling, on the crystal curvature for the first time with electrons and absolutely for the first time with a germanium crystal. 

\section{Theory of channeling, dechanneling and volume reflection}

For better understanding of this paper, we herewith provide a brief summary of orientational coherent phenomena in a bent crystal.

Planar channeling consists in the confinement of a charged particle trajectory by the electric field of crystal planes, when the particle transverse energy is smaller than the interplanar potential barrier (see Fig. \ref{Fig1}, where the model of Doyle-Terner potential \cite{DT} is applied), described as 
\begin{equation}
U_{eff}(x)=U(x)+pvx/R,
\label{1}
\end{equation}
where $U(x)$ is the interplanar potential for a straight crystal, $p$ and $v$ the particle momentum and velocity respectively, and $R$ the bending radius. Channeling may occur when the particle trajectory is nearly parallel to the crystal planes, more precisely for an incidence angle smaller than the Lindhard angle \cite{Lindhard} defined as
\begin{equation}
\theta_L=\sqrt{\frac{2 U_0}{pv}},
\label{2}
\end{equation}
$U_0$ being the potential well depth of a straight crystal. For 855 MeV electrons at MAMI channeled in the (111) planes one obtains the following values of the Lindhard angle: $\theta_{L,Si}=$ 232 $\mu $rad and $\theta_{L,Ge}=$ 274 $\mu $rad, for Si and Ge, respectively. 

If the crystal is bent, the well depth and the Lindhard angle decrease with $R$, going to zero for the critical radius:
\begin{equation}
R_{cr}=\frac{pv}{E_0},
\label{3}
\end{equation}
where $E_0$ is the maximal value of the interplanar electric field. For $R \leq R_{cr}$ channeling is forbidden. In case of 855 MeV electrons moving in the field of (111) planes, the critical radius is $R_{cr,Si} =1.59$ mm and $R_{cr,Ge}=1$ mm for Si and Ge, respectively.

Channeling proved to be an efficient way to steer positive particle beams, achieving deflection efficiencies larger than 80 \%. On the other hand, the maximal deflection efficiency for electrons recorded in the literature slightly exceeds 20 \% \cite{PRL2014} and is about 30 \% for negative pions \cite{SPS2013}. The main source of inefficiency is \textit{dechanneling}, which consists in particles escaping from the channeling mode caused by Coulomb scattering on atoms \cite{Biryukov,Baier,Tikhomirov}.The process of dechanneling is determined by the dechanneling length, which is the mean free path of a channeled particle before its transverse energy becomes larger than the potential barrier, thus escaping from the channeling condition (CH in Fig. \ref{Fig1}). Electron dechanneling lenght has already been experimentally measured with Si straight \cite{backe1,backe2,backe3} and bent \cite{PRL2014,EPJC2017,PRABSLAC,PRLSLAC} crystals. In contrast to positrons, electrons dechannel faster \cite{PRL2014,EPJC2017,PRABSLAC,PRLSLAC,Biryukov,Baier,Tikhomirov}. Indeed, negatively charged channeled particles oscillate around atomic planes, thereby being more subject to the strong scattering with lattice nuclei. 

The dechanneling length $L_{dech}$ is usually defined \cite{Biryukov} according to the dependence of the channeling fraction population $f_{ch}$ on the penetration depth $z$   
\begin{equation}
f_{ch}=A_0 \exp(-z/L_{dech}), 
\label{31}
\end{equation}
where $A_0$ is the normalizing factor. Both the dechanneling lenght and exponential channeling fraction decay were introduced by Kumakhov to describe the dechanneling process induced by electron scattering of non relativistic ions. This approach is well grounded for positively charged particles \cite{Biryukov} while even the recent inclusion of nuclear scattering \cite{Tikhomirov}, which is dominant for the dechanneling of negative particles, failed to extend quantitatively the same approach to this second case. Despite all this, the usage of the dechanneling length is justified by the practical use as a qualitative characteristical lenght for channeling, which depends on the crystalline material, thichkness and bending radius. Thereby, extrapolating $L_{dech}$ from experimental results is of interest for channeling application. The main diffuculty on the theoretical description of electron dechanneling process is the frequent strong changes in the negative particle transverse motion that leads to a limited applicability of a diffusion approach. These effects can anyway be taken into account in Monte Carlo simulations. For instance, the process of recapture under the channeling conditions of a dechanneled particle, the so-called rechanneling, was described well only using Monte Carlo simulations \cite{PRL2014}. However, in first approximation the exponential character of dechanneling (Eq. (\ref{31})) for negatively charged particles is maintained for highly-bent crystals, as shown by different experiments as well as Monte Carlo simulations \cite{PRL2014,PRABSLAC,PRLSLAC,SPS2013}, ensuring its usage for our experimental cases (see Sec. 4).

The transverse energy, $\epsilon$, of a charged particle moving in the field of bent crystal planes is
\begin{equation}
\epsilon = U_{eff}(x) + pv \theta_x^2 /2,
\label{32}
\end{equation}
being $x$ the transverse coordinate and $\theta_x=dx/dz$ the incidence angle w.r.t. the atomic plane, where $z$ is the longitudinal coordinate.
Since a particle cannot be captured under channeling mode if its transverse energy $\epsilon$ exceeds the potential well height, the channeling efficiency strongly depends on the crystal alignment as well as on the angular divergence of the incident beam. On the other hand, if an over-barrier particle moves towards the interplanar potential (as shown in Fig. 1), it can be reflected by a potential barrier, i.e. it experiences the so-called volume reflection. The typical angle of such reflection $\alpha$ is comparable with the Lindhard angle \cite{Maisheev}. 

One can calculate the maximal angle for VR integrating Eq. (\ref{32}) by varying $x$ from infinity to the reflection point $x_{ref}$ (see Fig. \ref{Fig1}) and \textit{vice versa}.
Moreover, by using (\ref{2}-\ref{3}) one can write the volume reflection angle \cite{Maisheev,Maisheev2} in a form independent of the beam energy:
\begin{eqnarray}
\begin{array}{l}
\frac{\alpha}{\theta_{L}} = \frac{E_0}{\eta \sqrt{U_0}}\int\limits_{x_0}^{x_c(\epsilon)} \left[\frac{1}{\sqrt{\epsilon-U(x)-\frac{E_0}{\eta}x}}- \frac{1}{\sqrt{\epsilon-U(x_c(\epsilon))-\frac{E_0}{\eta}x}} \right]dx,
\end{array}
\label{33}
\end{eqnarray}
where $\eta=R/R_{cr}$. This equation should be averaged on the transverse energy values, i.e. [$U_0$,$U_0+E_0 d_0/\eta$], where $d_0$ is the period of the interplanar potential $U(x)$. usTh one obtains:
\begin{eqnarray}
<\frac{\alpha}{\theta_{L}}> = \frac{\eta}{E_0 d_0}\int\limits_{U_0}^{U_0+\frac{E_0}{\eta}d_0} \frac{\alpha}{\theta_{L}} d\epsilon.
\label{35}
\end{eqnarray}
Formulae (\ref{33}-\ref{35}) do not take into account multiple scattering, though they provide a good estimate of the volume reflection peak position, as will be shown in Sec. IV. Since formulae (\ref{33}-\ref{35}) hold for both positively and negatively change particles, we will apply them for electrons.

Apart from being volume reflected, an over-barrier particle can lose its transverse energy while crossing the crystal plane near the reflection point because of incoherent scattering with lattice atoms and consequently be captured into the channeling mode (\textit{CH} in Fig. \ref{Fig1}). This effect is called volume capture \cite{VR}. 

Summarizing, all the coherent effects mentioned above, such as channeling and VR, are strongly dependent on the ratio between the initial transverse energy and the planar potential well depth, which depends on the bending radius, and hence on the incidence angle. In Sec. 4, a detailed and quantitative investigation of this dependence is presented, to obtain the optimal parameters of bent crystals for applications.

\section{Bent crystal manufacturing and experimental setup.}

A sample holder prototype aimed to bend the crystal with a remote controlled system was realized at the INFN-LNL lab in Legnaro, Italy. This innovative holder permitted to experimentally investigate important bent crystal parameters for application, such as the channeling and VR deflection angle and efficiency and the dechanneling length as a function of the curvature radius $R$, without manually re-bending the crystal. This allows a smooth increase of the curvature, performing it by numerical control, avoiding stress concentrations that causes the sample breaking during the mounting procedures in a normal fixed curvature sample holder. The sample holder is also equipped with a remotely controlled distortion correction system. If the sample curvature is not perfect, the channeling angle could vary along the beam dimension (torsion) causing a detrimental effect on efficiency \cite{IPAC2010}. The correction system is an additional degree of freedom allowing one to vary the torsion when the sample is into the chamber, to immediately check the effect on the efficiency. Furthermore, this innovative holder permits to re-bent the crystal without vacuum breaking, permitting a considerable speed up of the data taking to access a more extended set of data.

\begin{figure}
\resizebox{33mm}{!}{\includegraphics{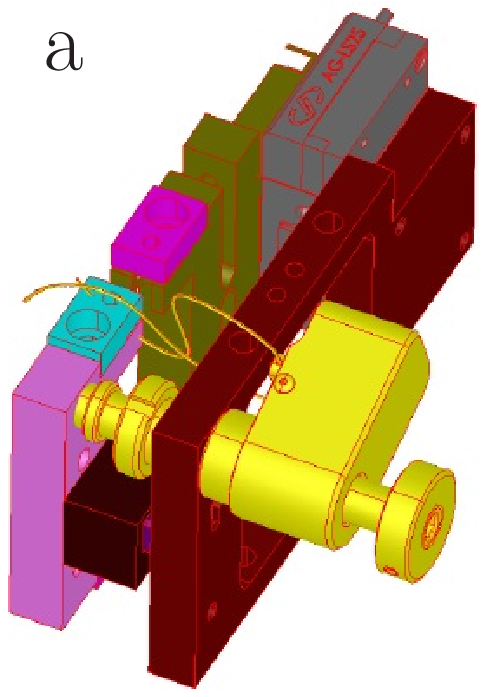}}
\resizebox{43mm}{!}{\includegraphics{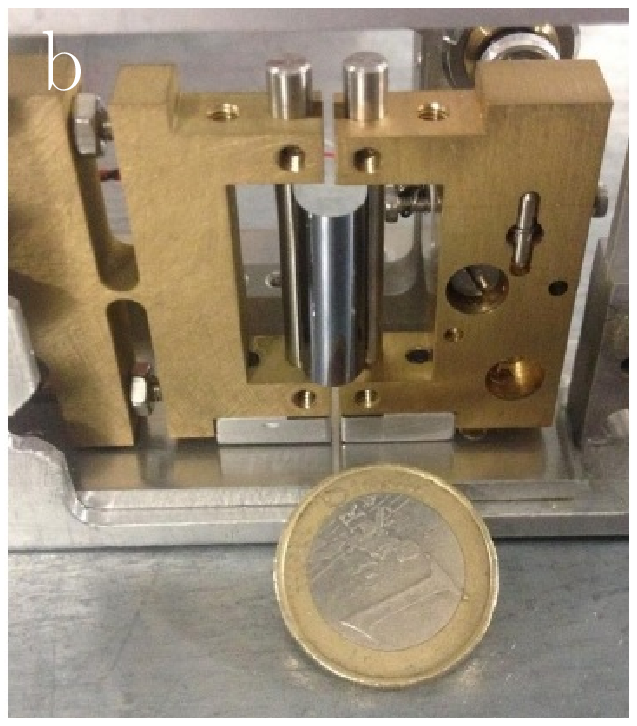}}\\
\resizebox{85mm}{!}{\includegraphics{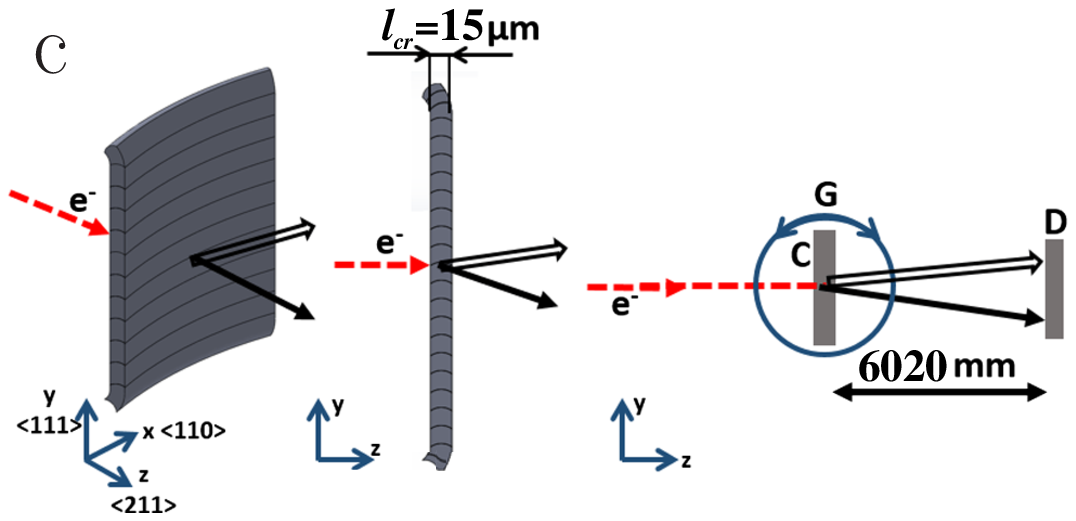}}
\caption{\label{Fig15} Piezo motor (a), dynamical holder (b) and bending of a silicon crystal (c). The dashed arrow indicates the incoming beam, impinging on the crystal mounted on a high-precision goniometer (G). The solid-black arrow indicates particles deflected under the channeling mode, while the solid-white arrow corresponds to VR
particles. The deflected beam impinges on the LYSO Screen (D).}
\end{figure}

Fig. \ref{Fig15}a shows the piezo motor step (grey) that translates a movable part (green) with respect to a fixed one (brown). It can be done with 400 nm steps. Two plugs are in-built in the two parts and translate one against the other by actioning the step motor. The plugs (steel cylinders) have two rabbets to secure the sample by a special gluing procedure. The plugs can rotate freely in brass holes and are supported by pins screws. When the plug gets closer due to translation caused by the motor, the sample is forced to bend (see Fig. 2b). To regulate the torsion a fine  movement is obtained by a never ending piezo driven screw (yellow) that pushes a rotating part (magenta) containing the plug hole. The angular resolution is better than 1 $\mu $rad. 
The system was calibrated in order to know the primary curvature radius as a function of the number of steps of the translator motor. The capability of the second motor to modify the crystal torsion was verified by means of high-resolution X-ray diffraction. 

The 15 $\mu $m long Si and Ge crystals samples were sequentially mounted onto the dynamical bending holder; the picture in Fig. \ref{Fig15}b shows the Si sample mounted on the holder. Such an ultralow length is essential for our experiment, because it must not exceed too much the dechanneling length as well as to reduce the multiple scattering angle. A bending moment supported the crystal at two opposite edges, leaving it free at the remaining edges. In this way, the crystal surface, which is parallel to the (211) planes, was bent along the (111) direction, obtaining a secondary bending of the (111) planes due to the quasimosaic effect \cite{QM}. The advantages of crystals exploiting the quasimosaic effect is represented by the possibility to manufacture ultra thin crystals large enough to completely intercept the beam. In addition, due to the shape of the potential well, the (111) bent planes (see Fig. \ref{Fig1}) are the most efficient for the deflection of negatively charged particles. The anticlastic effect could be a drawback of quasimosaic bent crystals but in the case of the present data, the large bending and small thickness guarantee a complete anticlastic suppression as demonstrate in \cite{Anticlastic}. This was checked by high resolution X-ray diffraction by measuring the (111) orientation at different positions along the y-axis. No anticlastic trend was evidenced with an exception of about 2mm close to the sample border. On the other hand, a residual variation of the (111) plane orientation of about $\pm$ 50 $\mu $rad inside the dimension of the measuring X-ray beam (100 $\mu $m) was evidenced. This is interpreted as a residual sample rippling induced by imperfections caused by the glueing procedure.

An experiment was carried out in the Hall B of the Mainzer Mikrotron (MAMI) with 855 MeV electrons. The experimental setup is the same as in \cite{Lietti} with the substitution of the Si microstrip detector to measure the beam profile after the interaction with the crystal with a LYSO Screen (see Fig. \ref{Fig15}c). The screen has a thickness of 200 $\mu $m and is inclined of 22.5 degrees toward the camera in the perpendicular direction with respect to the beam deflection plane and was placed downstream the crystal of 6020 mm. The crystal holder was mounted on a high-precision goniometer with 5 degrees of freedom. Translations along the x and y axes were used to geometrically align the crystal with the beam direction, while rotations around the x, y and z axes with an accuracy of 17.5, 30, and 50 $\mu$rad respectively, were used to achieve angular alignment of the crystal planes with the electron beam. The entire experimental setup was kept under vacuum to avoid multiple scattering of the beam by air.
The beam was focused through dedicated quadrupole lenses: the resulting beam size and angular divergence were 105 $\mu $m and 21 $\mu $rad along the vertical direction, which is the crystal bending direction. The beam divergence is smaller of the Lindard critical angle for channeling, which is about 220 $\mu $rad at 855 MeV. 
A schematic view of the experimental setup is shown in Figure 2(c): it allows to characterize, with very high precision, Si and Ge crystals in terms of both deflection efficiency and dechanneling length.


\begin{figure*}
\resizebox{85mm}{!}{\includegraphics{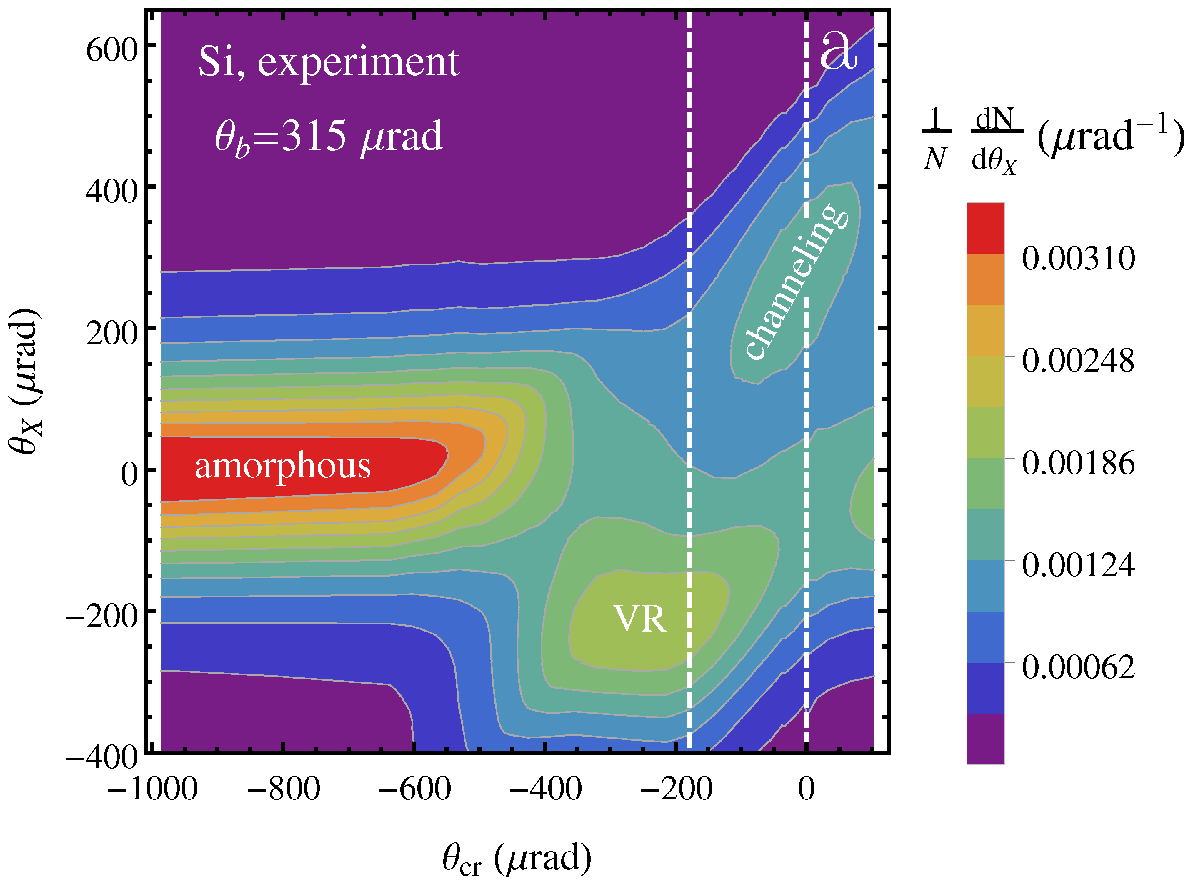}}
\resizebox{85mm}{!}{\includegraphics{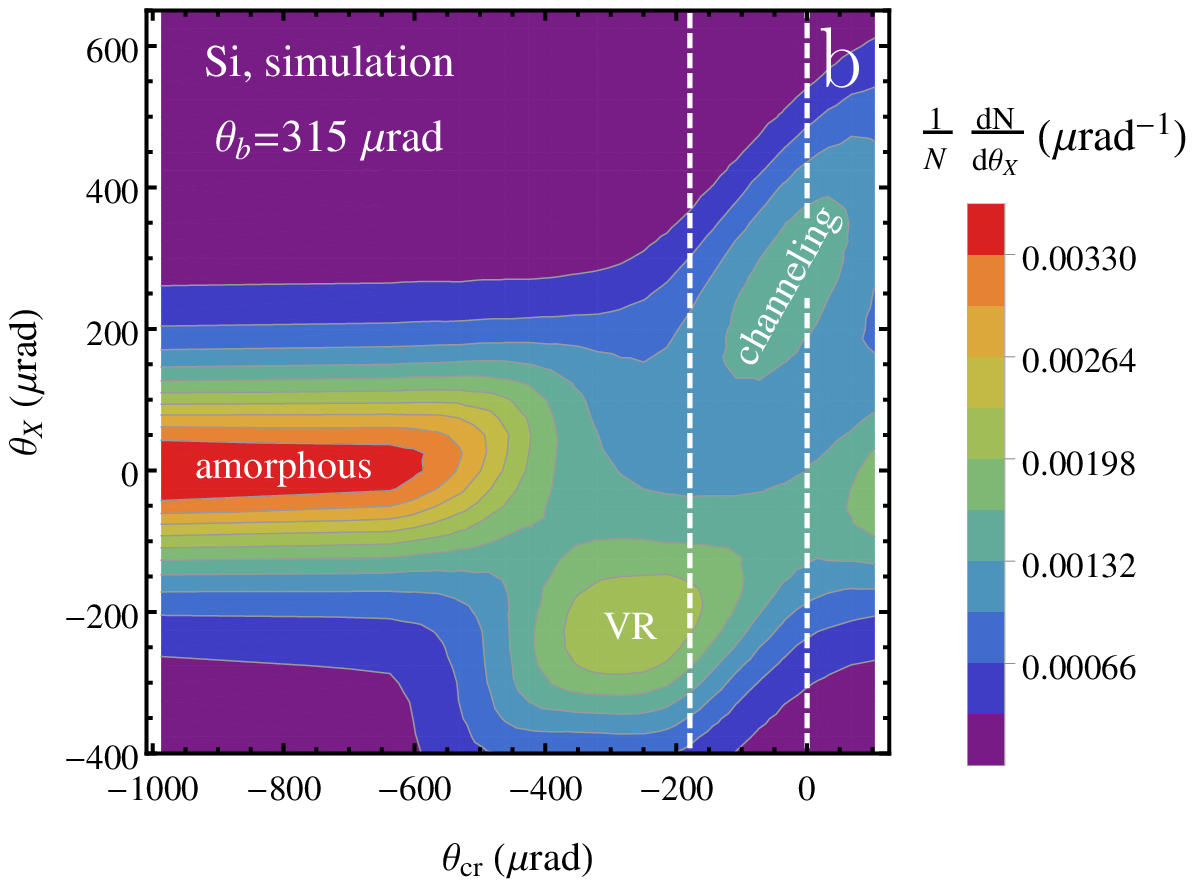}}\\
\resizebox{85mm}{!}{\includegraphics{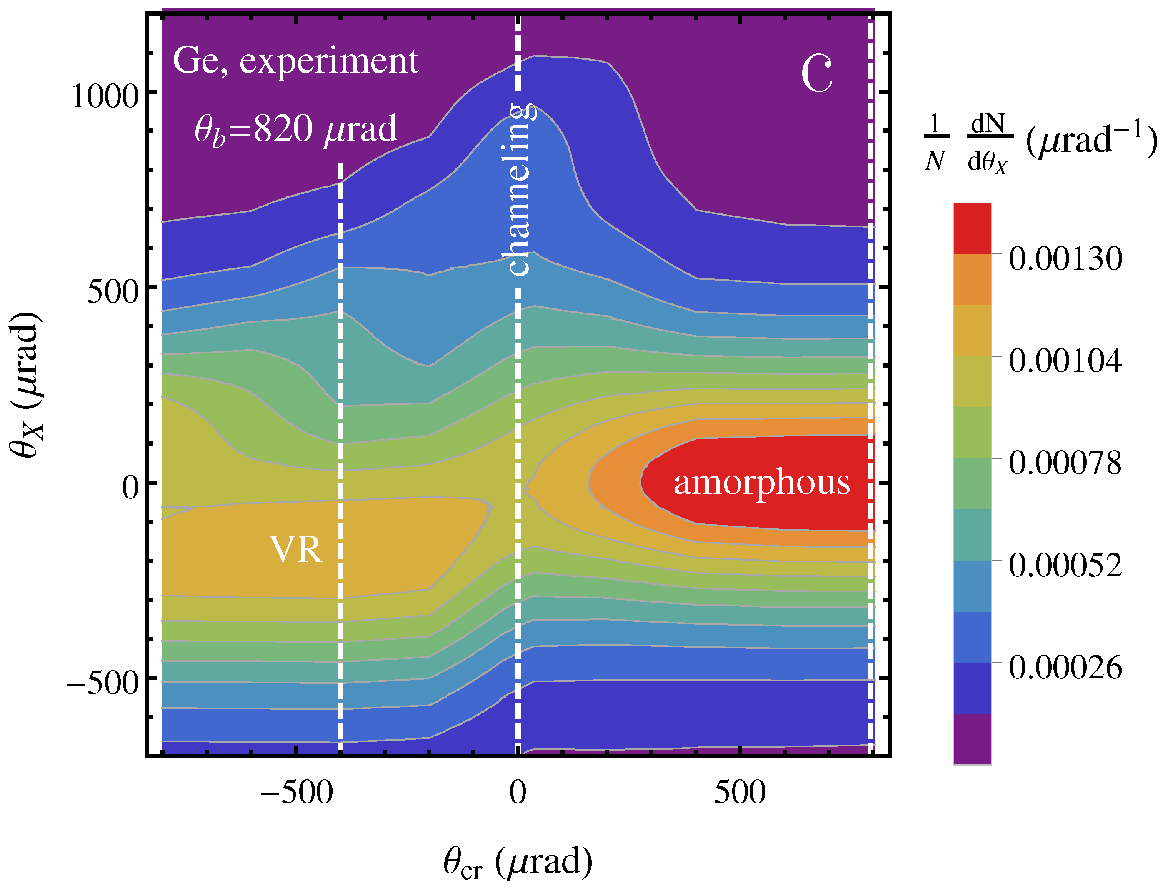}}
\resizebox{85mm}{!}{\includegraphics{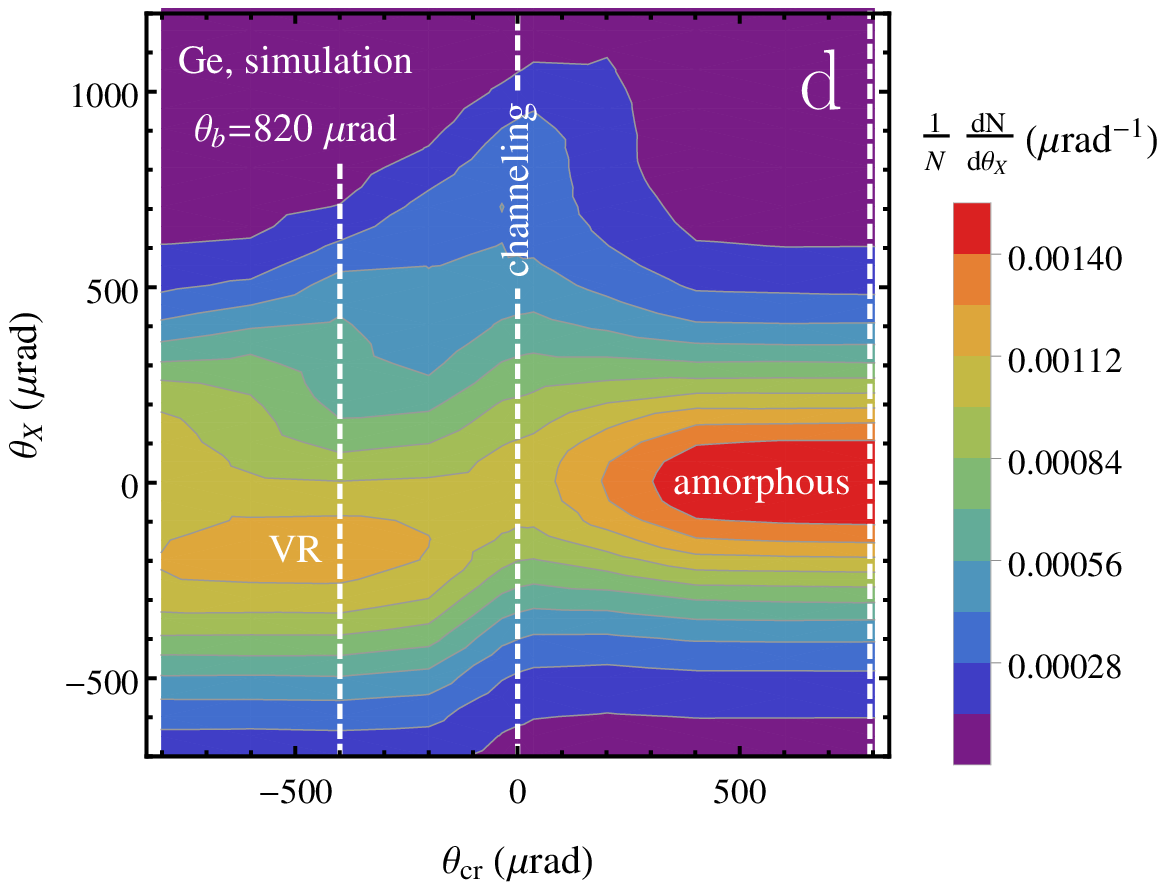}}
\caption{\label{Fig2} Experimental (a,c) and simulated (b,d) angular scans. Deflection angle vs. the crystal-to-beam orientation of Si (a,b) and Ge (c,d) crystals, bent at 315 $\mu $rad and 820 $\mu $rad respectively.}
\end{figure*}
\begin{figure*}
\resizebox{85mm}{!}{\includegraphics{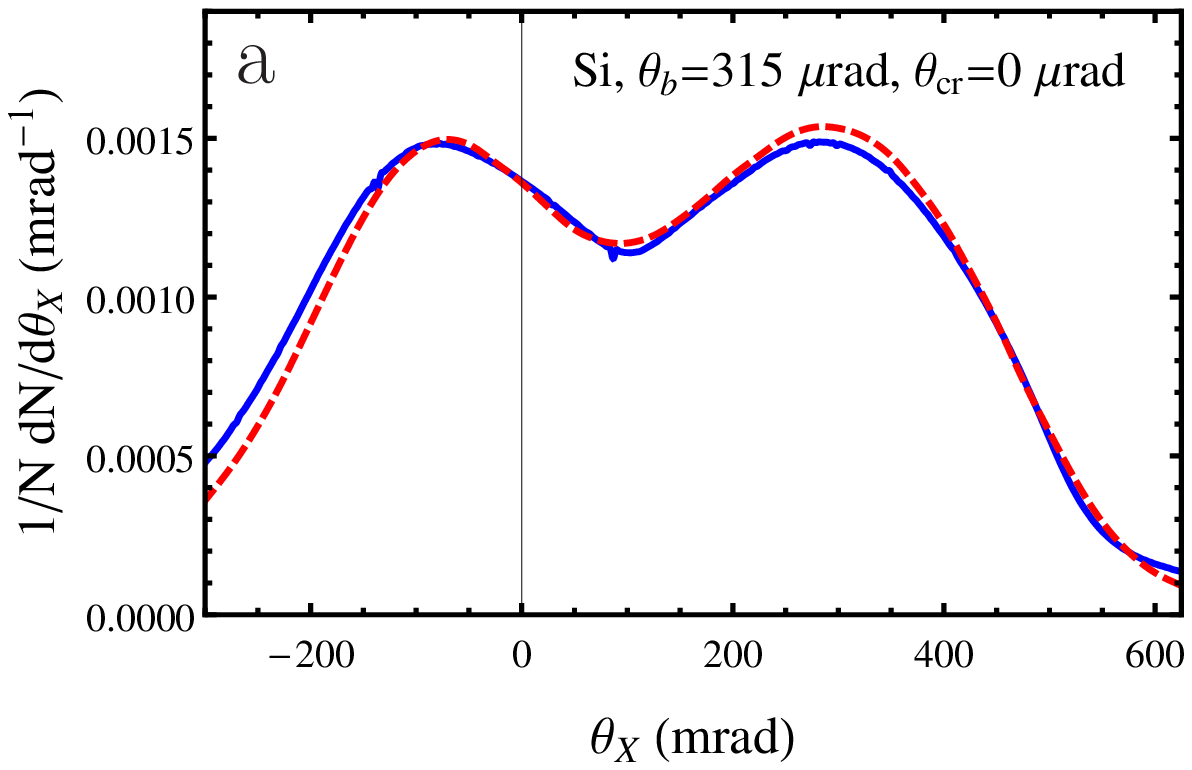}}
\resizebox{85mm}{!}{\includegraphics{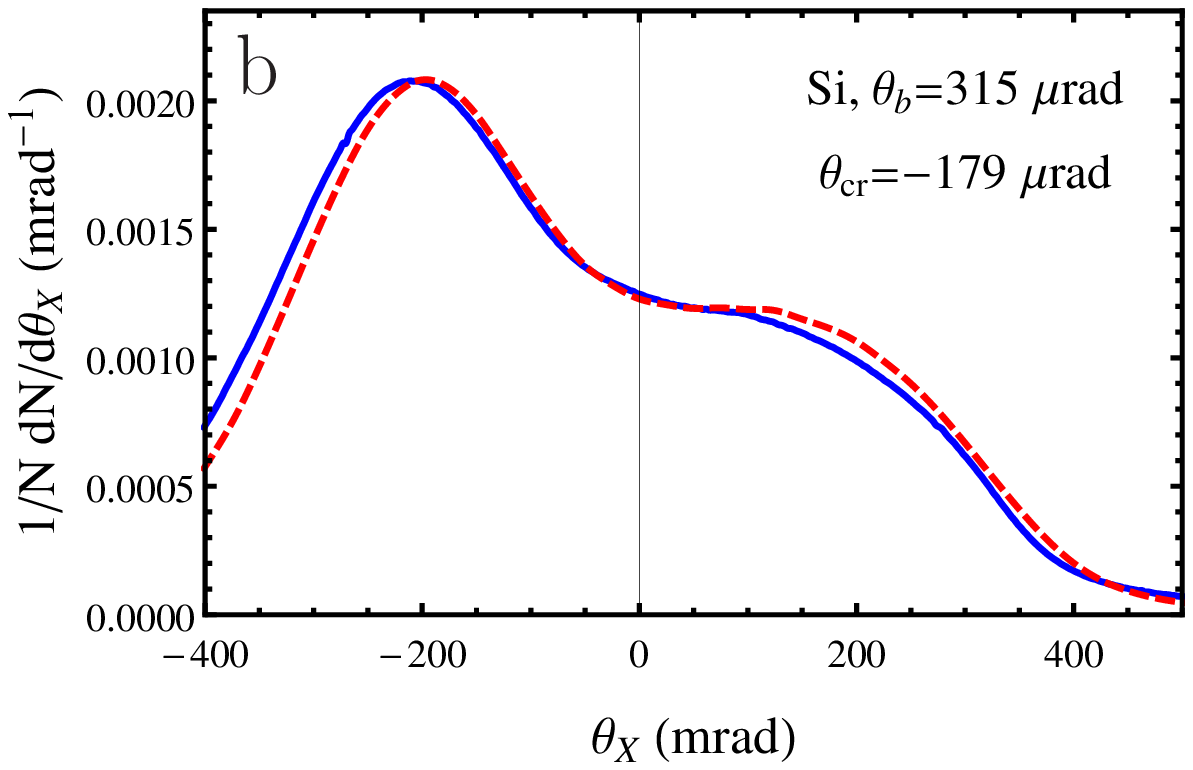}}\\
\resizebox{85mm}{!}{\includegraphics{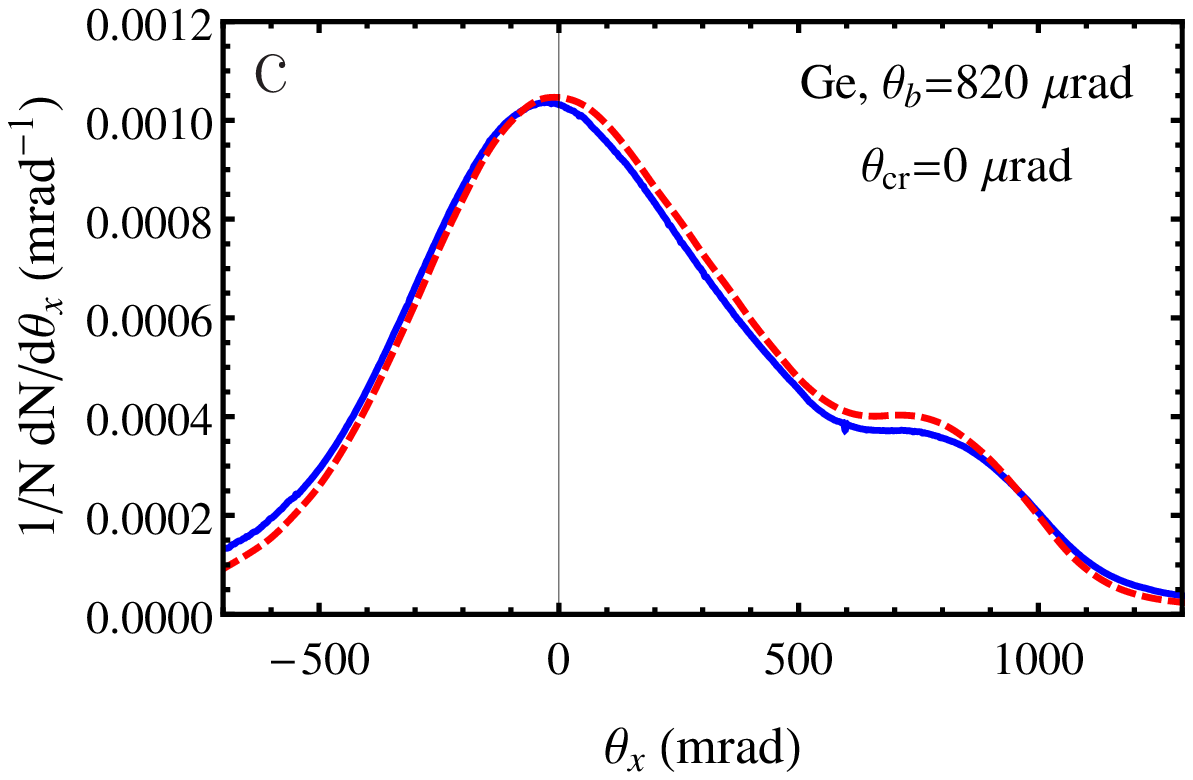}}
\resizebox{85mm}{!}{\includegraphics{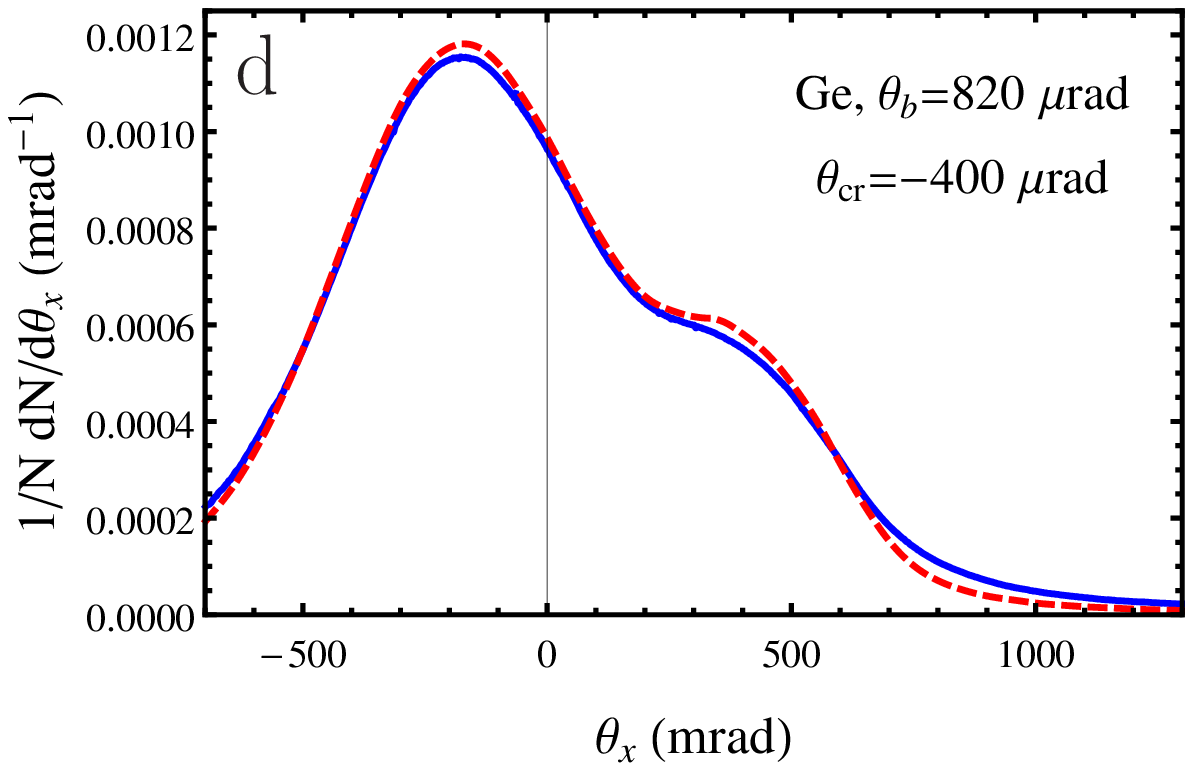}}
\caption{\label{Fig3} Experimental (solid) and simulated (dashed) distributions of deflected beam by Si (a,b) and Ge (c,d) crystals, bent at 315 $\mu $rad and 820 $\mu $rad respectively, for channeling (a,c) and volume reflection (b,d) crystal alignment. Experimental values represent the projection of the beam spot collected by the screen along the vertical direction (y) of bending.}
\end{figure*}

\section{Experimental results and analysis}

By exploiting the dynamical bending holder, we tested channeling and VR of 855 MeV electrons for 4 and 3 different curvatures in case of Si and Ge crystals, respectively. For each bending radius, $R$, we measured the distribution of the particles angles $\theta_{x}$ after the interaction with the crystal vs. crystal-to-beam orientation, $\theta_{cr}$, by rotating the goniometer around the ideal alignment with bent (111) planes. As an example, the experimental angular scan for the silicon crystal with a deflection angle of $\theta_b=315$ $\mu $rad and the germanium crystal of $\theta_b=820$ $\mu $rad are shown in Fig. \ref{Fig2}. Fig. \ref{Fig3} shows the beam deflection distributions with Si and Ge crystals oriented in channeling and in the middle of VR region, for the angular position highlighted by dashed lines in Fig. \ref{Fig2}.  These plots allow one to follow the transition between the main processes occurring while changing the crystal-to-beam orientation. In the angular distributions for channeling orientation ($\theta_{cr}=0$) the right peak represents the channeling mode as well as the left one is for the over-barrier particles. By decreasing $\theta_{cr}$ to about the middle of the range [$-\theta_{b}+\theta_L$;$-\theta_L$] one sets up the VR orientation. At the latter, the right peak represents volume captured particles. Finally, a crystal alignment beyond this range suppresses all the coherent effects, leading to the ``amorphous'' region where multiple scattering dominates. 

The crystal bending angle and alignment has been measured experimentally and verified by computer simulations using the CRYSTAL simulation code \cite{CRYSTAL1,CRYSTAL2}. This simulation code is developed for ab-initio Monte Carlo simulations of charged particle trajectories in an interplanar or interaxial potential in a crystalline medium (either bent or not) with both multiple and single scattering on nuclei and electrons. The method developed by the code has already been tested vs. experimental results in \cite{PRA2012,NIMB2013,PRL2013,PRL2014,PRL2017}. The perfect alignment with bent planes, i.e., $\theta_{cr}=0$, was experimentally determined by the highest intensity of the channeling peak, recorded during the angular scan. The channeling peak position also provided the crystal bending angle, verified by CRYSTAL simulations. The Monte Carlo simulations permitted to take into account the incoming angle distribution and sample rippling. 
The outcomes of the CRYSTAL code are displayed in Figs. \ref{Fig2} and \ref{Fig3} for comparison with experimental measurements. 

The analysis of the angular distributions was carried out through a fitting procedure based on the one presented in Ref. \cite{PRABSLAC,EPJC2017}. The fitting function represents the sum of the channeling part, described by gaussian:
\begin{eqnarray}
\begin{array}{l}
\frac{df_{ch}}{d \theta_X} = 
\frac{A_{ch}}{\sigma_{ch} \sqrt{2 \pi}} \exp \left(-\frac{(\theta_X-\theta_{ch})^2}{2 \sigma^2_{ch}} \right),
\label{36}
\end{array}
\end{eqnarray}
the volume reflection part, containing also a non-reflected overbarrier fraction and described by the sum of two gaussians:
\begin{eqnarray}
\begin{array}{l}
\frac{df_{VR}}{d \theta_X} = 
\frac{A_{VR}}{\sigma_{VR} \sqrt{2 \pi}} \exp \left(-\frac{(\theta_X-\theta_{VR})^2}{2 \sigma^2_{VR}} \right)+\\
\frac{1-A_{VR}}{r \sigma_{VR} \sqrt{2 \pi}} \exp \left(-\frac{(\theta_X-\theta_{VR})^2}{2 r^2 \sigma^2_{VR}} \right),
\label{37}
\end{array}
\end{eqnarray}
and the dechanneling part, being an exponential distribution, convolved with the first gaussian in (\ref{37}):
\begin{eqnarray}
\begin{array}{l}
\frac{df_{dech}}{d \theta_X} = 
\frac{A_{dech}}{2 \theta_{dech}} \exp \left(\frac{\sigma_{VR}^2}{2 \theta_{dech}^2}+\frac{\theta_{ch}-\theta_X}{\theta_{dech}}\right)\times \\ \left( \erf \left(\frac{\theta_{VR}-\theta_X+\frac{\sigma_{VR}^2}{\theta_{dech}}}{\sqrt{2} \sigma_{VR}}\right)-\erf \left(\frac{\theta_{ch}-\theta_X+\frac{\sigma_{VR}^2}{\theta_{dech}}}{\sqrt{2} \sigma_{VR}}\right)\right).
\label{38}
\end{array}
\end{eqnarray}

The total fitting function can be written as:
\begin{equation}
\frac{1}{N} \frac{dN}{d \theta_X} = \frac{df_{ch}}{d \theta_X}+B_{VR} \frac{df_{VR}}{d \theta_X}+\frac{df_{dech}}{d \theta_X}.
\label{39}
\end{equation}

In (\ref{36}-\ref{39}) $A_{ch}$, $A_{VR}$, $B_{VR}$, $A_{dech}$ and $r$ are the normalizing factors, $\theta_{ch}$, $\theta_{VR}$ and $\sigma_{ch}$, $\sigma_{VR}$ the mean angles and the standard deviations of corresponding gaussians respectively as well as $\theta_{dech}$ the ``dechanneling angle'', defining the dechanneling length, found from the angular distribution, as $L_{dech}=R \theta_{dech}$. The channeling efficiency is defined as the integral value of the gaussian fit of the channeling peak (\ref{36}), within $\pm 3 \sigma_{ch}$ around the channeling peak, namely $\eta_{ch} \approx 0.9973 A_{ch}$.

The fit procedure was carried out in two steps. First, Eq. (\ref{37}) was applied for the fit of the angular distribution of the crystal, aligned in amorphous direction. The values $A_{VR}$ and $r$, extracted in the first step were used in the fit (\ref{39}) \cite{PRABSLAC,EPJC2017}.

The main difference with the fitting procedure from \cite{PRABSLAC,EPJC2017} are the coefficients $A_{ch}$, $B_{VR}$, $A_{dech}$, treated independently. Though the increase of the free parameters number reduces the accuracy, it is necessary in this case for a correct description of initially overbarrier particles, as will be explained later in the text. 

In order to provide the most accurate simulation results as possible, the simulated channeling efficiency values were directly computed by using the CRYSTAL simulation code through the calculation of channeled (under-barrier) particles population. 
The deflection efficiency obtained through the fitting of simulated beam profiles is nearly the same calculated directly counting the number of under-barrier particles for $R/R_{cr} < 20$, determining the goodness of the fitting procedure (8-11) to estimate the deflection efficiency. The limitation of this procedure in the range $R/R_{cr} > 20$ is connected with the overlap of the channeling and over-barrier peaks in the deflected beam profile for too high bending radii as explained later in the text.

\begin{figure}
\resizebox{85mm}{!}{\includegraphics{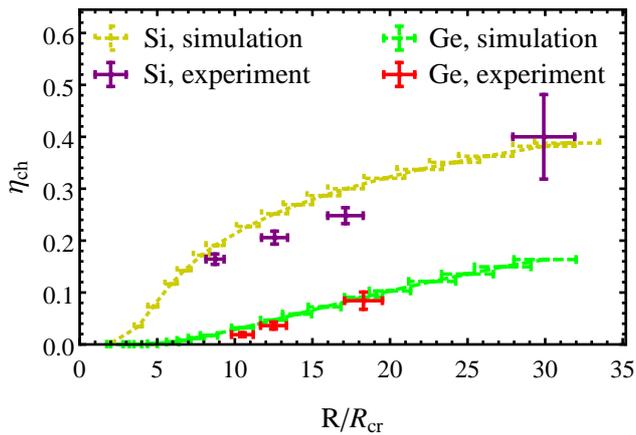}}
\caption{\label{Fig5} The experimental and simulated depedences of channeling efficiency of Si and Ge crystals at optimal channeling orientation on the ratio of the bending radius to its critical value.}
\end{figure}

The experimental results were critically compared to CRYSTAL simulations, highlighting a good agreement between them. The dependence of the channeling deflection efficiency on the ratio $R/R_{cr}$ is shown in Fig. \ref{Fig5} for both experimental and simulation results. The errors on the experimental efficiencies are due to the fitting error with an additional uncertainty connected with the normalization procedure. On the other hand, the x-error of simulated results is connected with the uncertainty of the crystal length, while the small y-error is due to statistics.
Table \ref{Table1} displays all values of the curvature radii, bending angles, $\theta_b$, and channeling efficiency used in the experiment; Table \ref{Table2} represents the same results obtained with simulations.  As expected, the dependence of channeling efficiency is monotonic \cite{EPJC2014}, since the potential well depth decreases while $R$ becomes smaller.

The experimental results highlighted a channeling efficiency larger than 35 \% for silicon in agreement with simulations.
Through the fitting procedure (8-11) it was not possible to extract the dechanneling length for silicon in the case $R/R_c>20$, while the channeling efficiency values was found with very large errors as explained later in the text. 
The experimental error is rather high for the curvature of $\theta_b=315$ $\mu $rad, because channeling and volume reflection peaks are very close, and it is difficult to distinguish the channeling fraction. By this reason, the angular distance between the channeling and volume reflection peaks is the main restriction of the fit (8-11). Nevertheless, it is clear from Fig. \ref{Fig3} upper left that high-efficiencies as those in this paper have never been achieved so far for electrons.

On the other hand, channeling efficiency for germanium achieves 8 \% at the lowest experimental bending angle. This is indeed the first evidence of negative beam deflection via channeling in a bent Ge crystal. 

Although, channeling efficiency for germanium is much lower than for silicon, this effect should not be attributed to the quality of the crystal, because both germanium and silicon crystals were manufactured through the same procedures leading to high performance of both crystals at much higher energy \cite{Ge1,Ge2,Ge3}. The only reason for such a difference owes to the influence of Coulomb scattering, which is about 2.2 times stronger for Ge than for Si. Indeed, this angle can be roughly estimated by multiple scattering formula \cite{PDG}:
\begin{equation}
\theta_{sc}=\frac{13.6 \text{MeV}}{pv} \sqrt{l_{cr}/X_{rad}}[1+0.038 \ln(l_{cr}/X_{rad})],
\label{4}
\end{equation}
where $X_{rad}$ is the radiation length, $l_{cr}$ the crystal length along the beam direction. By substituting the crystal parameters into (\ref{4}), one obtains the estimated multiple scattering angles for silicon and germanium crystals, being 130 $\mu $rad and 290 $\mu $rad, respectively. While the first value is 1.8 less than the Lindhard angle, the second one is of the same order. This fact explains our choice of ultra-thin crystals (15 $\mu $m), otherwise multiple scattering would cover all the coherent effects, leading to the impossibility to measure neither channeling or VR. 

\begin{table}[b]
\caption{\label{Table1}%
Measured Si and Ge crystal bending radia, angles and channeling efficiency}
\begin{tabular}{llllll}
\hline\noalign{\smallskip}
\textrm{Material}&
\textrm{$\theta_{b}$ ($\mu $rad)}&
\textrm{$\frac{R}{R_{cr}}$}&
\textrm{$\theta_{VR}$ ($\mu $rad)}&
\textrm{$\eta_{ch}$}\\
\noalign{\smallskip}\hline\noalign{\smallskip}
Si& $315$  & $29.9$ & $224$ & $0.40 \pm0.08 $\\
Si& $550$  & $17.1$ & $204$ & $0.248\pm0.016 $\\
Si& $750$  & $12.6$ & $194$ & $0.206\pm0.013 $\\
Si& $1080$ & $8.72$ & $183$ & $0.165\pm0.010 $\\
Ge& $820$  & $18.3$ & $172$ & $0.084\pm0.017 $\\
Ge& $1200$ & $12.5$ & $165$ & $0.036\pm0.007 $\\
Ge& $1430$ & $10.4$ & $162$ & $0.019\pm0.004 $\\
\noalign{\smallskip}\hline
\end{tabular}
\end{table}

\begin{figure}
\resizebox{85mm}{!}{\includegraphics{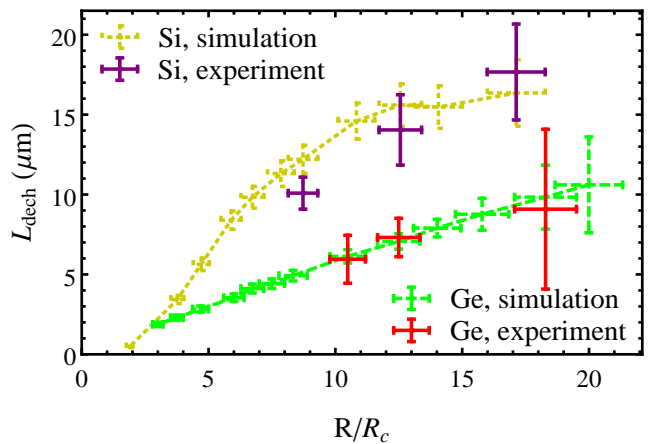}}
\caption{\label{Fig6} The experimental and simulated dependence of the dechanneling length on the ratio of the bending radius to its critical value at ideal channeling orientation. The dechanneling length for the first silicon curvature of $\theta_b=315$ $\mu $rad, was not extracted because channeling and volume reflection peaks were too close, thus making the fit (\ref{36}-\ref{39}) practically inapplicable.}
\end{figure}

\begin{figure*}
\resizebox{66.5mm}{!}{\includegraphics{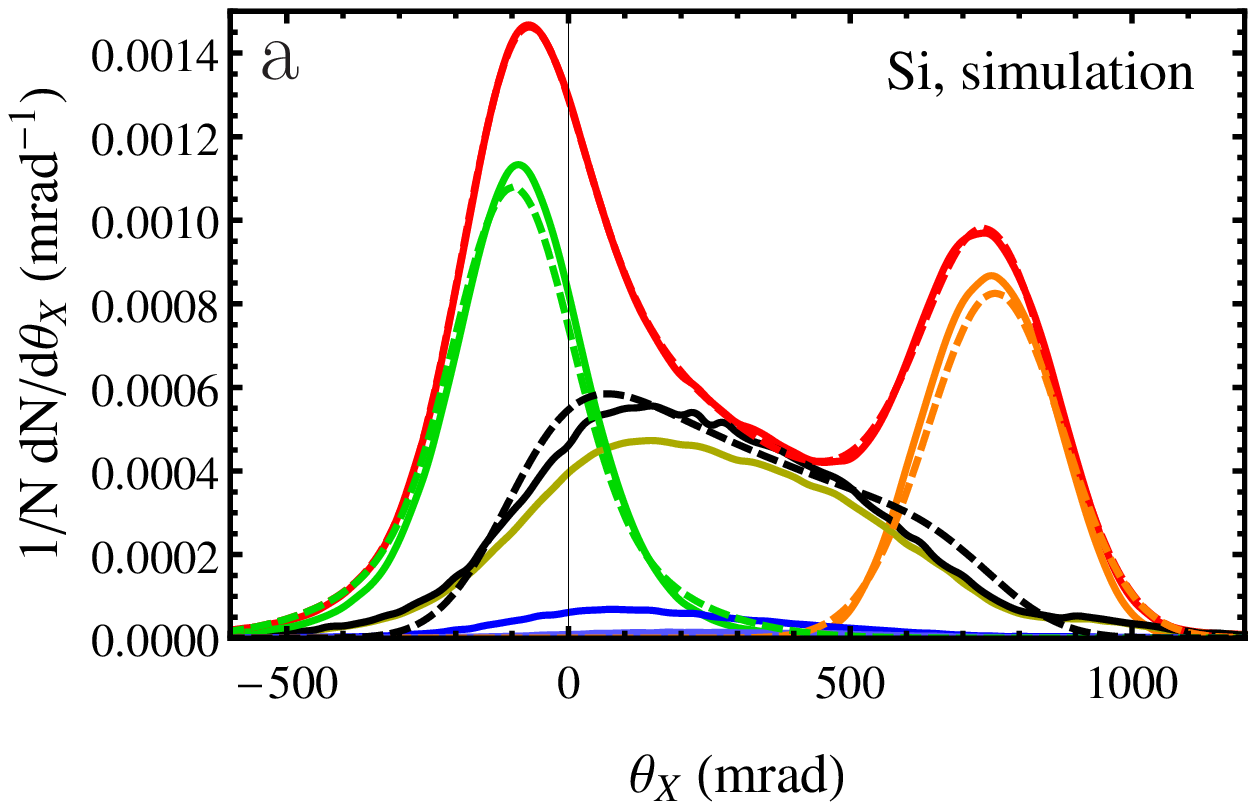}}
\resizebox{107mm}{!}{\includegraphics{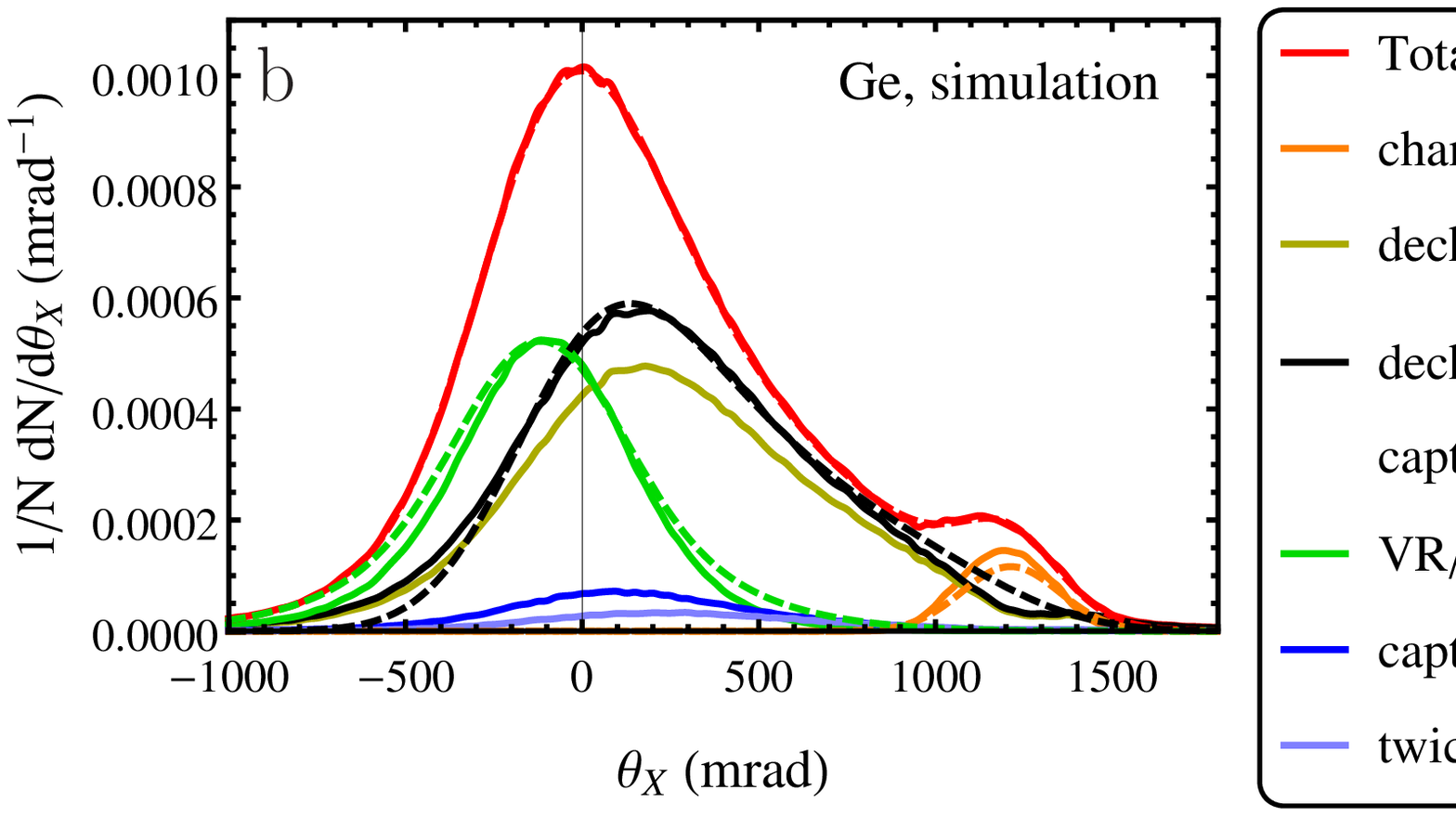}}
\caption{\label{Fig65} Simulated distributions of deflected beam by Si (left) and Ge (right) crystals, bent at 750 $\mu $rad and 1200 $ \mu $rad respectively, and the channeling, dechanneling, initially overbarrier, overbarrier, captured under and escaped the channeling mode one and two times (captured overbarrier) and the sum of captured overbarrier and dechanneling fractions. Dashed curves represent the total, channeling, overbarrier and dechanneling fractions, obtained by means of the fit (\ref{36}-\ref{39}).}
\end{figure*}
\begin{figure}
\resizebox{85mm}{!}{\includegraphics{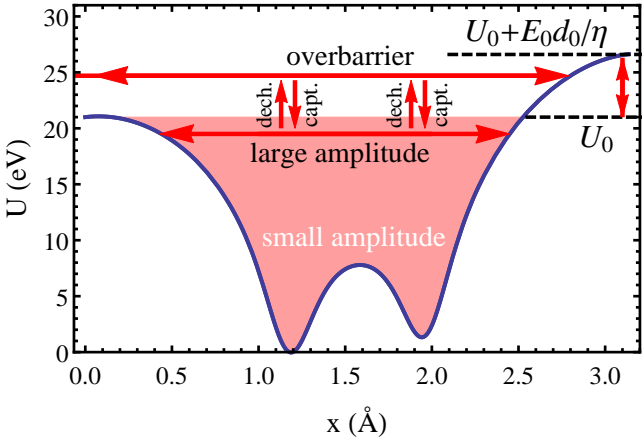}}
\caption{\label{Fig75}  Interplanar potential of (111) Si for maximal bending radius (4.76 cm) used in the experiment. Overbarrier marks initially overbarrier fraction of particles that can be captured into the channeling mode or rechaneled (rech.) and dechanneled (capt.) once or several times.
}
\end{figure}
\begin{figure}
\resizebox{85mm}{!}{\includegraphics{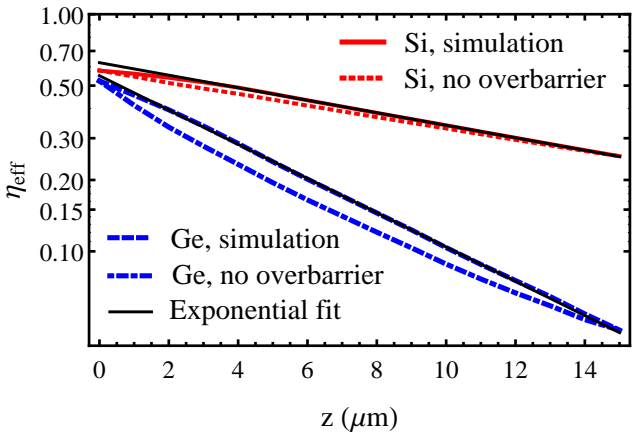}}
\caption{\label{Fig7} Exponential fit of the depedence of the channeling efficiency in the penetration depth in Si and Ge crystals with the same parameters as in Fig. \ref{Fig65} and the same dependences without the overbarrier particles, captured under and escaped the channeling mode.}
\end{figure}
\begin{figure*}
\resizebox{73mm}{!}{\includegraphics{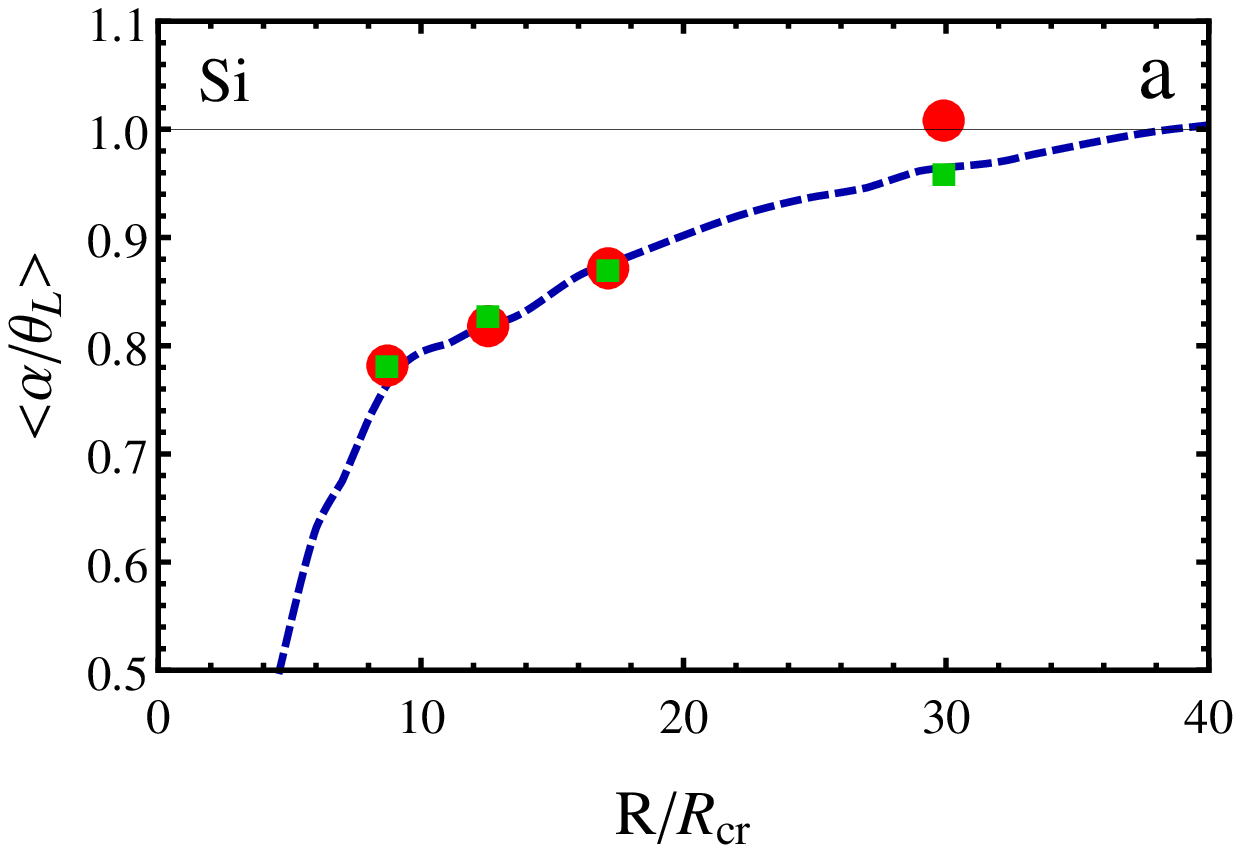}}
\resizebox{100mm}{!}{\includegraphics{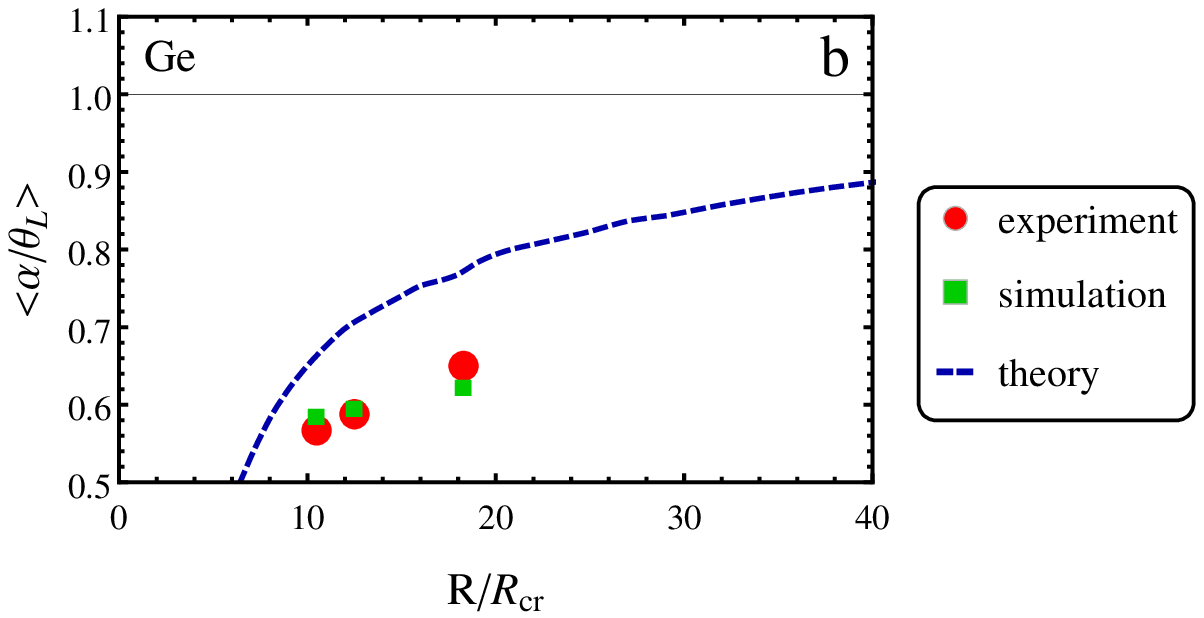}}
\caption{\label{Fig8} Theoretical, experimental and simulated dependence of the maximal value of volume reflection peak position w.r.t. the Lindhard angle on the bending radius w.r.t. its critical value for both Si and Ge (111) bent crystals.}
\end{figure*}

To complete the analysis on channeling, one should evaluate the main parameter that determines the steering capability of a crystal through the dechanneling length. Such parameter has been extracted by using the fit (\ref{36}-\ref{39}) of both experimental and simulated deflection distributions. The dependences of extracted dechanneling length on the ratio of bending radius and critical radius for both silicon and germanium are shown in Fig. \ref{Fig6}. The corresponding experimental and simulation values are listed in Table \ref{Table3}. 

As for channeling efficiency, the measured dechanneling length depends monotonically on the crystal radius in agreement with simulations. The silicon dechanneling length is comparable with the length of the crystal, while for germanium being at least 1.5--3 times less. This fact explains the difference in channeling efficiency between the two materials, being due to the different multiple scattering contribution for different atomic number Z.

The present data demonstrate that negative particles steering efficiency is mainly regulated by dechanneling length and not by the channeling well depth that would have benefit Germanium. It is worth to note that this is a peculiar feature of negative particles since for positive ones, the influence of the potential well depth dominates the dechanneling process, and once Ge and Si efficiency for short crystal are compared, Ge performances prevail on Si \cite{Ge1,Ge2,Ge3}. This insight into the channeling performances by changing the atomic number suggests that low scattering materials such as diamond could be an interesting candidate to be investigated to improve the steering efficiency.

\begin{table}[b]

\caption{\label{Table2}%
Simulated Si and Ge crystal bending radia, angles and channeling efficiency}
\begin{tabular}{llllll}
\hline\noalign{\smallskip}
\textrm{Material}&
\textrm{$\theta_{b}$ ($\mu $rad)}&

\textrm{$\frac{R}{R_{cr}}$}&
\textrm{$\theta_{VR}$ ($\mu $rad)}&
\textrm{$\eta_{ch}$}\\
\noalign{\smallskip}\hline\noalign{\smallskip}
Si& $315$  & $29.9$ & $235$ & $0.3818\pm0.0004 $  \\
Si& $550$  & $17.1$ & $203$ & $0.3000\pm0.0004 $  \\
Si& $750$  & $12.6$ & $190$ & $0.2519\pm0.0003 $  \\
Si& $1080$ & $8.72$ & $182$ & $0.1907\pm0.0003 $  \\
Ge& $820$  & $18.3$ & $178$ & $0.0909\pm0.0002 $  \\
Ge& $1200$ & $12.5$ & $161$ & $0.0468\pm0.0002 $\\
Ge& $1430$ & $10.4$ & $156$ & $0.0320\pm0.0002$\\
\noalign{\smallskip}\hline
\end{tabular}
\end{table}

Given the good agreement between experiments and simulations, we may exploit the latter to investigate deeply the dechanneling process. In particular, one may separate the different contributions on the dechanneling distribution (10), between the VR and channeling peaks.
In fact, already in \cite{PRL2014} it was demonstrated that the rechanneling process (capture under channeling of dechanneled particles, see section 2) may have a strong influence on the dechanneling length. Here we investigate also the contribution of overbarrier particles to the dechanneling distribution. Fig. \ref{Fig65} displays different fractions in the angular distributions obtained directly from simulations (solid), namely channeling, dechanneling with taking into account rechanneling as well as the volume reflection/overbarrier fraction for both silicon (a) and germanium (b). For comparison the same fractions were extracted from the angular distributions by using the fit (\ref{36}-\ref{39}) (dashed).
\begin{table}[b]

\caption{\label{Table3}%
Experimental ($L_{dech\text{ Exp}}$) and simulated (both from distribution ($L_{dech\text{ Sim}}$) and directly from the dependence of channeling efficiency on the penetration depth ($L_{dech\text{ DSim}}$) as well as from the same dependence excluding overbarrier particles captured under the channeling mode and then dechanneled ($L'_{dech\text{ DSim}}$) Si and Ge dechanneling lengths. All dechanneling lengths are measured in $\mu$m.}
\begin{tabular}{llllll}
\hline\noalign{\smallskip}
\textrm{ }&

\textrm{$\frac{R}{R_{cr}}$}&
\textrm{$L_{dech\text{ Exp}}$}&
\textrm{$L_{dech\text{ Sim}}$}&
\textrm{$L_{dech\text{ DSim}}$}&
\textrm{$L'_{dech\text{ DSim}}$}\\
\noalign{\smallskip}\hline\noalign{\smallskip}
Si& $17.1$ & $17.7\pm3.0$ & $16.4\pm2.1$ & $18.96\pm0.05$& $21.14\pm0.10$\\
Si& $12.6$ & $14.0\pm2.2$ & $15.6\pm1.4$ & $16.48\pm0.05$& $18.05\pm0.07$\\
Si& $8.72$ & $10.1\pm1.0$ & $12.2\pm0.9$ & $13.62\pm0.05$& $14.73\pm0.06$\\
Ge& $18.3$ & $9 \pm 5$    & $10\pm 2$    & $7.97\pm0.07$ & $8.95\pm0.26$\\
Ge& $12.5$ & $7.3\pm1.2$  & $7.1\pm0.5$  & $6.02\pm0.03$ & $6.46\pm0.11$\\
Ge& $10.4$ & $5.9\pm1.5$  & $6.1\pm0.4$  & $5.29\pm0.03$ & $5.58\pm0.09$\\
\noalign{\smallskip}\hline
\end{tabular}
\end{table}

Fig. \ref{Fig65} highlights a contribution of initially overbarrier particles, that can be captured and dechanneled from the channeling mode several times (marked in Fig. \ref{Fig65} as rechanneled ovebarrier). In a bent crystal, the contribution of overbarrier particle cannot be eliminated, even considering a parallel beam. Indeed, due to the asymmetry in the potential barrier introduced by the bending, particles approaching with zero transverse kinetic energy to the right potential barrier are reflected to the left, gaining a non-zero transverse kinetic energy. A sketch of initially overbarrier particles is depicted in Fig. \ref{Fig75}; such particles follow the bent crystal planes and can be deflected to a considerable angle. 

The solid black line in Fig. \ref{Fig65} represents the contribution of dechanneled and rechanneled particles with the contribution of captured overbarrier particles. If compared with the dotted solid black line, representing the result of the fit (\ref{36}-\ref{39}), it is clear that from the experimental deflection distribution is not possible to extract a dechanneling length correspondent to only initially channeled particles. By this reason the fit \cite{PRABSLAC,EPJC2017} was modified to (\ref{36}-\ref{39}).

To highlight deeply the contribution of captured overbarrier particles, channeling efficiency was also simulated in a dependence on the penetration depth $z$, taking (solid) and not taking (dashed) into account this contribution for both Si (red) and Ge (blue) crystals as shown in Fig. (Fig. \ref{Fig7}). By using an exponential fit (\ref{31}) the values of dechanneling length were extracted (see Table \ref{Table3}, $L_{dech}\text{ }_{DSim}$ and $L'_{dech}\text{ }_{DSim}$, for the cases with and without captured overbarriers, respectively). The values $L_{dech}$ $_{DSim}$ differ from the extracted ones from experimental and simulated angular distributions, no more than on $\sim$ 1--2 $\mu$m, lying usually within the frame of the error. 

On the contrary, the simulated dechanneling length without the contribution of captured overbarrier particles, $L'_{dech}\text{ }_{DSim}$, (see Table \ref{Table3}), exceed $L_{dech}\text{ }_{DSim}$ by 5--10\%. In other words, the initially overbarrier particles decrease the total dechanneling length by several percent because can usually be captured slightly below the potential well barrier, as shown in Fig. \ref{Fig75}. Since these large amplitute particles dechannel faster, the dechanneling length of these particles is lower than for the stable
ones. Consequently, the capture of initially overbarrier particles reduce the total measured dechanneling length. Furthermore, even if these values were obtained in the same way as $L_{dech}\text{ }_{DSim}$, the depedence of $L'_{dech}\text{ }_{DSim}$ on the penetration depth evidently differ from the exponent function. Indeed, one has to remember that the dechanneling of negative particles is mainly due to strong scattering with nuclei that has an intrinsic non-slow diffusive nature \cite{Tikhomirov}.

The relative difference between $L'_{dech}\text{ }_{DSim}$ and \text{ }\text{ }\text{ } $L_{dech}\text{ }_{DSim}$ increases with crystal radius rise. This is explained by decreasing of the difference $E_0 d_0/\eta$ ($\eta=R/R_c$, see section 2) between the right and left potential barriers. Consequently such overbarrier particles are closer to the potential boundary. This means that such particles will remain near the potential barrier for a longer distance due to low transverse velocities, having the influence on the total dechannelling length also for a longer distance at higher radius values. Therefore, initially ovebarrier particles, captured under channeling mode and then dechanneled, can make a several percent contribution into dechanneling length value.

Finally, we also investigated the other mechanism of beam deflection, i.e., the VR. In particular we studied VR deflection angle vs. the curvature radius, while comparing to the maximal angle expected from the theory (\ref{33}-\ref{35}). 

In order to verify the theoretical dependence of the maximal angle of VR on $R$ (see Eqs. \ref{33}-\ref{35}), we used the experimental and simulated values for the modules of a maximal VR angle (determined by gaussian fit) comparing them in Fig. \ref{Fig8}. For channeling deflection angle, the agreement between theory and both experimental and simulated results is very good for silicon, while being worse for germanium. This fact is explained again by the contribution of multiple scattering of non-volume reflected particles, allocated around 0 angle (see Figs. \ref{Fig2},\ref{Fig3}), that shifts the volume reflection peak center towards 0 on the angular distribution. Multiple scattering has a much stronger influence for germanium, for which its r.m.s. angle is 2.2 times higher than for silicon. By this reason the measured maximum VR for silicon of 235 $\mu $rad is about $\theta_L$ angle for silicon (in agreement with previous experiments \cite{PRL2014,PRLSLAC}), while it is only $0.6\theta_L$ for germanium, being equal to 178 $\mu $rad.

\section{Conclusions}

An experiment on beam steering of 855 MeV electrons by using 15 $\mu $m bent silicon and germanium crystals has been carried out at the Mainzer Mikrotron. Through the exploitation of an innovative piezo-actuated mechanical bender, it was possible to test planar channeling and volume reflection for several radii of curvature.

Experimental results, in agreement with Monte Carlo simulation, demonstrated that maximum channeling efficiency were about 40\% and 8 \% for silicon and germanium, respectively. The difference between these two materials has to be ascribed to the higher atomic number $Z$ for Ge, which results in a higher Coulomb scattering contribution, causing stronger dechanneling. Indeed, we also measured the main parameter of planar channeling, i.e., the dechanneling length, which resulted to be close to the crystal length for Si, but 2 times shorter for the Ge crystal at the largest bending radius. In particular, the usage of a Si crystal with the length comparable to the dechanneling length permitted an unprecedented level of steering efficiency for an electron beam.

On the other hand, it is important to remark that any measurements of a negatively charged beam steering in a germanium bent crystal at the energies lower than hundreds of GeV have never been done before, due to the lack of properly designed crystals, i.e. with a length of the order of the dechanneling length. Therefore, the evidence of beam steering of sub-GeV electrons in a Ge crystal was demonstrated for the first time.

We also highlighted the influence of initially non-channeled particles on the dechanneling processes, which causes a reduction of the dechanneling length in case the crystal thicknesses are comparable with the dechanneling length. 


Finally, we investigated dependence of the ratio between the volume reflection angle and the Lindard angle vs. the $R/R_c$ (see Eq. (\ref{33}-\ref{35})), demonstrating that it does not depend on the energy, being very useful to make prediction at different energies.

The presented results, in particular the studies of beam efficiency and dechanneling length vs. the crystal curvature and atomic number, are of interest for application, such as generation of e.m. radiation in higher Z-materials bent and periodically bent crystals. Given the good agreement with Monte Carlo simulation, one may also think to apply the presented approach to extrapolate information on charged particle steering at higher energies, for instance to investigate the possibility of crystal-based collimation/extraction at current and future electrons accelerators.

\begin{acknowledgements}
We acknowledge partial support of the INFN-AXIAL experiment and by the European Commission through the PEARL Project within the H2020-MSCA-RISE-2015 call, GA 690991. We also acknowledge the CINECA award under the ISCRA initiative for the availability of high performance computing resources and support. E. Bagli, L. Bandiera and A. Mazzolari recognize the partial support of FP7-IDEAS-ERC CRYSBEAM project GA n. 615089. We acknowledge Professor H. Backe for fruitful discussions. We acknowledge M.Rampazzo, A. Pitacco and A. Minarello for technical assistance in dynamic holder realization.
\end{acknowledgements}


\begin{thebibliography}{00}
\bibitem{U70} A.G. Afonin et al., Phys. Rev. ST Accel. Beams \textbf{15}, 081001 (1-9) (2012)
\bibitem{U702} A.G. Afonin et al.,  JETP Letters \textbf{84} No. 7, 37276 (2006)
\bibitem{Tevatron} R. Carrigan et al., Phys. Rev. ST Accel. Beams \textbf{5}, 043501 (2002)
\bibitem{Tevatron2} N.V. Mokhov et al., Intern. J. of Mod. Phys. A \textbf{25}, Suppl. 1, 9875 (2010)
\bibitem{UA9} W. Scandale et al. Phys. Rev. ST: Accel. Beams \textbf{11}, 063501 (1-10) (2008)
\bibitem{UA92} W. Scandale et al., Phys. Lett. B \textbf{680}, 129-132 (2009)
\bibitem{LHC} W. Scandale et al., Phys. Lett. B \textbf{758}, 129-133 (2016)
\bibitem{Tsyganov} E. N. Tsyganov, Fermilab TM-682 (1976)
\bibitem{Lindhard} J. Lindhard, Mat. Fys. Medd. Dan. Vid. Selsk. 34 No. 14, 64 p.(1965)
\bibitem{VR} A. M. Taratin and S. A. Vorobiev, Nucl. Instrum. Methods Phys. Res., Sect. B 26, 512 (1987)
\bibitem{PRA2012} V. Guidi,  L. Bandiera, V. Tikhomirov, Phys. Rev. A 86, 042903 (2012)
\bibitem{NIMB2013} L. Bandiera et al. Nucl. Instr. and Meth. in Phys. Res. B 309, 135-140 (2013)
\bibitem{PRL2013} L. Bandiera et al. Phys. Rev. Lett. 111, 255502 (2013)	
\bibitem{PRL2015} L. Bandiera et al., Phys. Rev. Lett. 115, 025504 (2015)
\bibitem{PRL2014} A. Mazzolari et al., Phys. Rev. Lett. 112, 135503 (2014)
\bibitem{EPJC2017}E. Bagli et al. Eur. Phys. J. C 77:71 (2017)
\bibitem{PRABSLAC} T.N. Wistisen et al. Phys. Rev. Acc. and Beams 19, 071001 (2016)
\bibitem{PRLSLAC} U. Wienands et al. Phys. Rev. Lett. 114, 074801 (2015)
\bibitem{PS2011} S. Hasan et al. Nucl. Instr. and Meth. in Phys. Res. B 269, 612-621 (2011)
\bibitem{SPS2009ax} W. Scandale et al. Phys. Lett. B 680, 301-304 (2009)
\bibitem{SPS2009} W. Scandale et al. Phys. Lett. B 681, 233-236 (2009)
\bibitem{ILC} T. Behnke et al. arXiv:1306.6327 (2013)
\bibitem{CLIC} R. Toma\'{a}s Phys. Rev. ST -- Acc. and Beams 13, 014801 (2010)
\bibitem{RREPS2013} L. Bandiera et al. Journal of Physics: Conference Series 517, 012043 (2014)
\bibitem{SERYI} A. Seryi et al. Nucl. Instr. and Meth. in Phys. Res. A 623, 23 (2010)
\bibitem{PS2016} L. Bandiera et al. Proc. of Science (ICHEP2016), 069 (2016)
\bibitem{Ge0} C. Biino et al., Phys. Rev. B 403, 163 (1997)
\bibitem{Ge1} D. De Salvador et al. Appl. Phys. Lett. 98, 234102 (2011)
\bibitem{Ge2} D. De Salvador et al. AIP Conf. Proc. 1530, 103-110 (2013)
\bibitem{Ge3} D. De Salvador et al. Appl. Phys. Lett. 114, 154902 (2013)
\bibitem{DT} S.L. Dudarev et al., Surface Science 330, 86-100 (1995)
\bibitem{SPS2013} W. Scandale et al. Phys. Lett. B 719, 70 (2013)
\bibitem{Biryukov} V. Biryukov, Y. Chesnokov, and V. Kotov, Crystal Channeling and its Application at High-Energy Accelerators (Springer-Verlag, Berlin, 1997).
\bibitem{Baier} V. Baier, V. Katkov, and V. Strakhovenko, Electromagnetic
Processes at High Energies in Oriented Single Crystals (World Scientific, Singapore, 1998).
\bibitem{Tikhomirov} V.V. Tikhomirov, Eur. Phys. J. C 77:483 (2017)
\bibitem{backe1} H. Backe et al., Nucl. Inst. Meth. Phys. Res. B  266 (2008) 3835–3851
\bibitem{backe2} H. Backe et al., Nucl. Inst. Meth. Phys. Res. B 309 (2013) 37
\bibitem{backe3} H. Backe, and W. Lauth, Nucl. Instr. Meth. Phys. Res. B 355 (2015) 24-29
\bibitem{Maisheev} V.A. Maisheev, Phys. Rev. STAB 10, 084701 (2007)
\bibitem{Maisheev2} S. Bellucci et al., Phys. Rev. STAB 18, 114701 (2015)
\bibitem{IPAC2010} E. Bagli et al. Proc. of IPAC'10, THPEC080, 4243 (2010)
\bibitem{QM} R. Camattari, V. Guidi, V. Bellucci, A. Mazzolari, J. Appl. Cryst. 107, 064102-1--5 (2015) 10.1107/S1600576715009875
\bibitem{Anticlastic} Guidi et al. J. Phys. D: Appl. Phys. 42, 182005 (2009)
\bibitem{Lietti} D. Lietti et al. Rev. Sci. Instr. 86, 045102 (2015) 
\bibitem{CRYSTAL1} A.I. Sytov, V.V. Tikhomirov, Nucl. Instr. and Meth. in Phys. Res. B 355, 383-386 (2015)
\bibitem{CRYSTAL2} A.I. Sytov, Vestnik of the Bel. St. Univ. Series 1(2), 48-52 (2014)
\bibitem{PRL2017} T.N. Wistisen et al. Phys. Rev. Lett. 119, 024801 (2017)
\bibitem{EPJC2014} E. Bagli et al., Eur. Phys. J. C 74, 2740 (2016)
\bibitem{PDG} C. Patrignani et al. (Particle Data Group), Chin. Phys. C 40, 100001 (2016)

\end{thebibliography}
\end{document}